# Multicolor surface photometry of brightest cluster galaxies*.

## II. Enlarging the sample and improving the analysis

Stefano Andreon[1], Bianca Garilli[2], Dario Maccagni[2]

[1] Observatoire Midi-Pyrénées, URA 285, 14, Avenue Edouard Belin, F-31400 Toulouse, France
[2] Istituto di Fisica Cosmica del CNR, Via Bassini 15, I-20133 Milano, Italia

**Abstract.** Intrinsic color profiles of a sample of nine Brightest Cluster Galaxies (BCGs) are recovered from the observed color profiles by evaluating spurious gradients introduced by errors in the determination of the sky levels and by different seeing conditions between the observations. Isophote shapes and surface brightness profiles are presented for the four newly observed BCGs. Three out of nine BCGs show color gradients of the order of 0.10 mag per decade in radius. Five BCGs do not possess any color gradient larger than 0.01 mag per decade in radius. We do not see any correlation between the presence (or the sign) of a color gradient and the BCG morphology (slope and shape of its surface brightness profile), or the cluster richness. This argues against a strong and recent influence of the environment on the BCG morphology. The slope of the BCG surface brightness profile is correlated to the cluster richness, posing a constraint on the initial conditions, or on the physical mechanism that is responsible for the present BCG morphology. Finally, only one BCG can be a cD despite visual classification as such of all studied BCGs.

## 1. Introduction

The principal aim of many studies of Brightest Cluster Galaxies (BCGs) is to understand the processes leading to the formation of these giant galaxies, their evolution and relation with their environment. Moreover, these galaxies have been investigated to assess their reliability as distance indicators in the determination of the motion of the local group (Lauer & Postman 1994) and in cosmological tests, for instance in the Tolman test (Tolman 1930; see e.g. Sandage & Perelmuter 1991 for a recent application).

Surface brightness, color profiles and isophote shapes of these galaxies can give us a great deal of information on their properties (see Andreon et al. 1992, hereafter Paper I, and Mackie, Visvanathan & Carter 1990, hereafter MVC), but unfortunately the estimates of these quantities are strongly affected by the brightness of the sky and seeing conditions.

Surface brightness and color profiles of galaxies at large galaxy radii, i.e. at low galaxy brightness, are strongly affected by even a small error in the determination of the sky brightness (de Vaucouleurs & Capaccioli 1979; Capaccioli & de Vaucouleurs 1983). The measurement of the intrinsic surface brightness becomes more and more difficult as the redshift increases because of the cosmological dimming and the K correction. Moreover, the inner regions of the surface brightness profiles are affected by seeing to radii of 3-5 times the FWHM of the point spread function (Franx, Illingworth & Heckman, 1989). Consequently, the useful range for surface brightness and color profiles for galaxies at $z > 0.1$ is a decade in radius.

The shapes of galaxy isophotes are affected by poor seeing and nonuniform CCD response. The deviations of the isophote shape from perfect ellipses are usually small ($\sim 0.5\%$) for BCGs (MVC and Paper I), and often as large as their errors. It is thus important to accurately measure these deviations *and* compute their errors in order to understand the statistical significance of the computed deviations.

We devote this paper to enlarging the sample of BCGs with well studied isophote shapes, surface brightness and color profiles. Due to the importance of color profiles of BCGs, we concentrated on obtaining reliable color profiles for the largest possible radial range of the galaxies. The useful range of the color profile has been extended at large radii without using the value of the sky level, and at small galaxy radii by measuring the effect on the color profiles of the different seeing conditions between the exposures in the different filters. We measure the deviations of the BCG isophote shapes from perfect ellipses, and we assess their significance by comparing their values measured in different images. This analysis is done for 4 newly observed BCGs and, partially, for the five ones presented in Paper I. The larger number of BCGs and the better quality of the analysis allow us to compare the properties of BCGs of different morphologies and/or in different environments.

**Table 1.** The sample

| Cluster | B-M type | R | $z$ | ACO notes |
|---------|----------|---|-----|-----------|
| (1) | (2) | (3) | (4) | (5) |
| Abell 2335 | I* | 2 | $0.17^a$ | f.c. |
| Abell 2405 | I | 0 | $0.12^a$ | h.c. |
| Abell 3109 | I | 0 | 0.0673 | h.c., cD |
| S 84 | I-II | 0 | 0.11 | f.c., cD? |

f.c. and h.c. mean "faint corona" and "has corona" respectively.
$^a$ Estimated following Scaramella et al. (1991).
* Probably of intemediate B-M type.

## 2. The new data

### 2.1. Observations

We selected a sample of BCGs in clusters of Bautz-Morgan type I or I-II listed in Abell, Corwin and Olowin (1989, hereafter ACO). The main characteristics of the four BCGs studied in this paper are given in Table 1. Careful inspection of an ESO SERC film copy shows that the cluster A2335 is of intermediate Bautz-Morgan type and not a B-M type I. We were obliged to estimate the redshift of two clusters from Scaramella et al. (1991). Their redshifts have an uncertainty $\Delta z/z = 0.2$, which affects the determination of some photometric parameters, as discussed later.

The observations were carried out in September 1989 at the Cassegrain focus ($f/8.46$) of the 1.5 m Danish Telescope at La Silla equipped with an RCA Hi-Res CCD with 680x1024 pixels. The angular scale is 0.23 arcsec pixel$^{-1}$, resulting in a field of view of 2'.5x4'.0. As in Paper I, the images were taken through the Bessell $B,V$ (Bessell 1976) and the Gunn $i$ (Wade et al. 1979) filters.

Table 2 is the journal of the observations: column (1) gives the UT time of the observation; columns (2) and (3) the object observed and the filter used; column (4) gives the exposure time and column (5) the image seeing (FWHM).

### 2.2. Data reduction and analysis

The bulk of the data analysis does not differ from that presented in Paper I. The following minor changes apply:
- dark current was negligible, and was not subtracted from images;
- the CCD images of this run present a series of columns that are a bit hot or cold whit respect the adjacent ones. This bad column pattern, also called charge skimming, was removed in the way presented in Andreon (1993a);
- the sky in the $B$ and $V$ images appears flat: the dispersion of the modes of the sky levels measured in a series of boxes of 20x20 pixels is of the order of 0.2 to 0.5 % of the sky value. Moreover, coating defects are not appreciable in these images. Flat-fielding the images with our $B$ and $V$ dome flats does not improve the flatness of the sky and make its noise worse. For this reason, $B$ and $V$ images were not flat-fielded. $i$ images present intensity gradients and, for this reason, were flat-fielded.

**Table 2.** Summary of CCD observations

| Time | Object | Filter | Exp. time | Seeing |
|------|--------|--------|-----------|--------|
|      |        |        | sec       | arcsec |
| (1)  | (2)    | (3)    | (4)       | (5)    |
| 00.58.44 | Feige 284-9 | V | 180 | |
| 01.04.25 |  | B | 180 | |
| 01.11.02 |  | i | 90 | |
| 01.40.40 | Abell 2335 | V | 1200 | 1.15 |
| 02.03.44 |  | B | 1800 | 1.45 |
| 02.36.56 |  | i | 900 | 1.31 |
| 03.01.05 | E8-39S | i | 5 | |
| 03.05.20 |  | V | 10 | |
| 03.09.14 |  | B | 10 | |
| 03.24.54 | Abell 2405 | V | 1200 | 1.27 |
| 04.22.42 |  | i | 900 | 1.20 |
| 05.38.00 | E1-49V | V | 15 | |
| 05.43.04 |  | B | 20 | |
| 05.46.29 |  | i | 10 | |
| 06.31.41 | S84 | V | 1200 | 1.27 |
| 06.56.15 |  | B | 2400 | 1.61 |
| 07.39.49 |  | i | 900 | 1.27 |
| 08.42.11 | Feige 762-8 | V | 180 | |
| 08.47.47 |  | B | 180 | |
| 08.54.03 |  | i | 120 | |
| 09.07.52 | Abell 3109 | V | 900 | 1.31 |
| 09.26.41 |  | i | 600 | 1.15 |
| 09.40.09 |  | B | 900 | 1.38 |

### 2.3. Calibrations

The photometric calibration was obtained by observing Graham's stars E1-49P and E8-39S (Graham 1982) and the Feige sequence F284-9 and F762-8 (Stobie, Sagar & Gilmore 1985). Magnitudes and colors of the calibration stars were transformed into the proper photometric system as in Paper I. In the first part of the night, zero points computed from different calibration stars differ at most by 0.02 mag; the $B$ and $V$ ones computed from the sequence F762-8, observed at the end of the night, differ by 0.10 mag from the other ones. No $i$ zero point is computed from the observations of Feige 762-8 since there are no published useful colors of this sequence to be converted into the $i$ filter.

Since the observations of all BCGs, but A3109, are preceded and followed by the observation of a calibration star, the adopted zero-point is interpolated from the values computed from adjacent calibration observations. The zero-point of the $B$ and $V$ exposures of A3109 are computed from the sequence F762-8 whereas the $i$ zero-point is computed from the observation of the star E1-49V.

### 2.4. Sky determination

Sky levels were sampled as in Paper I, i.e. by measuring the counts in a series of 20x20 pixel boxes in areas away from the BCG and free from contaminating objects or chip defects. The dispersion of the mode of the sky level is, at worst, 0.5%. Much of this dispersion seems to be introduced by unresolved objects. The brightness of these objects is too small (0.1 to 0.3% of the sky brightness) to affect the isophote shape of the studied BCGs since these are obtained at relative high brightness (10 to 15% of the sky level)[1], but they are bright enough to affect the brightness profiles at 1 % of the sky level. To limit the effect of these unresolved objects, a sigma-clipping procedure that removes from 10 to 30 % of the brighter pixels on each isophote has been used to compute surface brightness profiles of BCGs. The effect of possible residuals after the removal of the brighter pixels should be lower for color profiles (at least if the apparent positions of the unresolved objects are the same in the two exposures), being color profiles differential measures.

Surface brightness profiles were computed out to galaxy radii where the surface brightness reaches 1 % of the sky level.

## 3. Improving the analysis

### 3.1. Isophote shapes of the new BCGs

By definition, the measure of the error of any quantity is given by the dispersion of the values measured in repeated identical physical observations. In order to measure the errors of the deviations of a galaxy isophote from the elliptical shape, we would need at least two images (in the same filter) of that galaxy. Unfortunately, as is common in astronomical observations, we have only one image in each filter and therefore we cannot compute the errors. Nevertheless, we can associate an error estimate to our data under the assumption that the deviations from the elliptical shape are the same in the three filters. We can then use the dispersion of the values measured in the three filters as a measure of the error. In this way we slightly overestimate the errors.

We think that this approach yields more realistic results than, for instance, the comparison of the values measured for nearby but not independent radii or the assumption of a relation between the errors of the deviations and the other shape parameters. We found that Jedrzejewski's (1987) approach, commonly used in the literature and also by us in Paper I, gives deviation errors that are 3 time larger than our measured dispersion. Since we overestimate the true error, this factor is a lower limit.

### 3.2. Determination of the color profiles

Different seeing conditions and errors in the determination of the *true* value of the sky level introduce spurious color gradients. We are able to disentangle spurious color gradients from intrinsic ones by comparing the observed color profile to the expected one obtained by taking into account the effects of seeing and errors in the determination of the sky value.

First of all, we developed a method for estimating the galaxy color profile that does not need any knowledge of the sky level and consequently is not affected by it. The method has been sketched in Andreon (1993b), whereas an extensive and exhaustive description can be found in Sparks & Jørgensen (1993) who found it independently, and to which we refer for details. Let us consider the function $I_{\lambda_1}(I_{\lambda_2})$ that represents the measured galaxy intensities in the filter $\lambda_1$ at the galaxy radius where the measured galaxy intensity in the other filter is $I_{\lambda_2}$. A color gradient due to an error in the determination of the sky level produces a linear shift of this function in the $I_{\lambda_1}$ vs $I_{\lambda_2}$ plane whereas an intrinsic color gradient produces a change of its (local) slope in the same plane. The two effects being different, a color gradient intrinsic to the galaxy can be discriminated from a color gradient produced by an incorrect choice of the sky level. Equivalently, in the *color vs. radius* plane, more usual for astronomers, among all the possible color profiles there is a family of curves generated by incorrect choices of the sky level. The latter differ in shape from the profiles due to intrinsic color gradients. We choose to compare our color profiles to the curve of this family which matches it in at least one point. This normalisation point is chosen at a galaxy brightness 4.5 mag arcsec$^{-2}$ fainter than the sky level to maximize the range over which the profiles can be compared. At fainter brightnesses the galaxy flux becomes too strongly affected by Poissonian noise and truncation errors to be useful.

We now consider the spurious color gradients due to different seeing conditions between the exposures. Seeing conditions have been measured from the images of stars extracted from the same frames. For each filter, we have convolved a model of the galaxy (a de Vaucouleurs' law with effective radius computed from the part of the galaxy profile unaffected by the seeing) with a model of seeing (an optical transfer function with $\eta = 5/3$ ; Woolf 1982) having the same FWHM of the star profile and constructed the expected color profile[2].

To understand the influence of our assumption of a unique effective radius for the whole radial range, we compared the expected color gradients of two galaxies which have very different effective radii. For radii larger than 1 arcsec, and for galaxies having effective radii similar to the ones observed, computed color profiles of galaxies having effective radii $a$ and $2a$ differ at most by only 0.02 mag arcsec$^{-2}$ when seeing conditions between exposures differ by 0.4 arcsec FWHM (more than the largest difference in our observations). Consequently, simulated galaxies whose surfaces brightness profiles are very different in the center (two de Vaucouleurs' laws having effective radii $a$ and $2a$ and same brightness at the $a$ radius differ by 1.32 mag arcsec$^{-2}$) have color profiles due to different seeing conditions not too much different. This gives us the possibility of parameterizing the unknown surface brightness profile of the galaxy in the region strongly affected by the seeing with a de Vaucouleurs' law having the same effective radius as the one fitted at large radii. Clearly, smaller deviations from de Vaucouleurs' law of the real surface brightness profiles of the studied galaxies will give smaller effects.

Moreover we also tested the assumption of a specific form for the seeing profile: changing $\eta$ from the theoretical value 5/3 to the value 1.5, found by Thomsen & Baum (1987) to be more appropriate to represent a real PSF, all but one color profile remain unchanged. The instrumental V-i color profile of A3670, the more distant BCG in the sample and observed in worse and more dissimilar seeing conditions, becomes more peaked in the center by 0.1 mag arcsec$^{-2}$.

Finally, for each color index and for each galaxy the color profiles due the two effects have been added to give the ex-pected color profile to which we compare the observations over all the studied range.

---

[1] In any case, the apparent dimensions of these unresolved objects being much smaller than the observed galaxies, they affect only high frequency coefficients of the development of the isophote shape in sine and cosine series (e.g. $a10$, $b10$, $a11$ ...) and not the studied ones ($a3$, $a4$, $b3$, $b4$).

[2] We assume that the effective radius of the galaxy is the same in the different filters, otherwise the galaxy would possess a color gradient by definition.

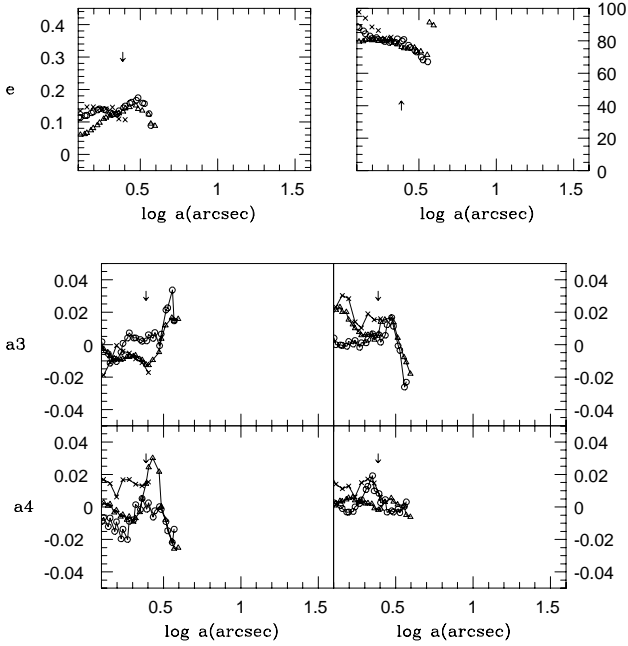

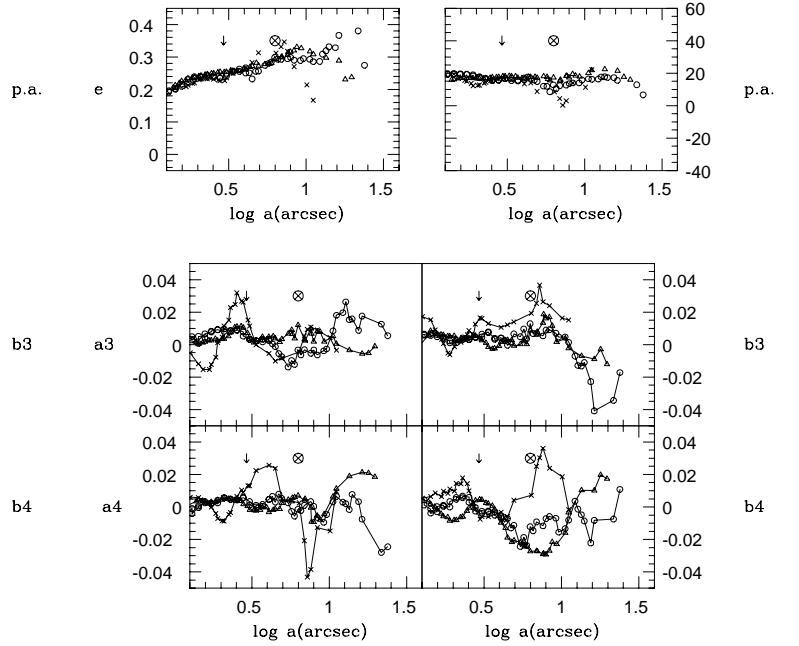

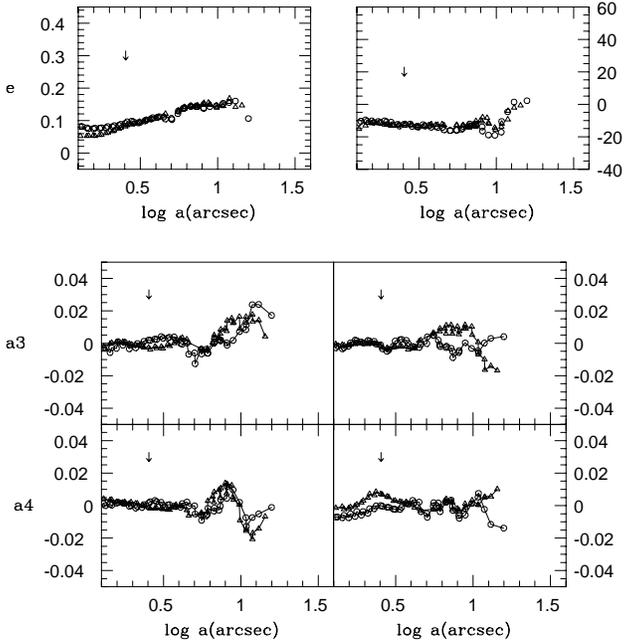

**Fig. 2.** Next page. Surface brightness and color profiles of the 4 BCGs of our new sample as a function of the logarithm of the semi-major axis $a$. The plotted lines represent the expected color profile due to different seeing conditions and errors in the determination of the sky value. In each filter, the fainter plotted surface brightness profile point is 5 mag arcsec$^{-2}$ fainter than the sky brightness.

**Fig. 1.** Ellipticity, position angle, fourier decomposition of the isophote deviations from perfect ellipses of A2335, A2405 and A3109 as a function of the logarithm of the semi-major axis $a$. Arrows indicate twice the FWHM of the image seeing. A circled cross at $\log(r) \sim 0.8$ in the A3109 plot shows the position of a nearby object. Crosses, triangles and circles mark values derived from the B, V and $i$ images respectively. For the sake of clarity the profiles a3, b3, a4 and b4 are smoothed by a 5x5, 3x3 and 5x5 top hat filter for the B, V and $i$ filters respectively.

## 4. Results

### 4.1. Isophote shapes

Isophote fitting has been performed using the STSDAS (Space Telescope Scientific Data Analysis System) isophote package (version 1.2). Details on the algorithm can be found in Jedrzejewski (1987). Objects (galaxies, stars, defects and cosmic rays) superposed to the BCG image have been masked. At large galaxy radii, the images have been rebinned to improve the S/N ratio. Figure 1 shows the ellipticity, position angle, and sine (b3 and b4) and cosine (a3 and a4) terms of the fitting function of the isophotes of all but one (S84) BCGs of our sample. The BCG S84 is so round that the matrix in the harmonic fit is singular, and does not allow an isophotal analysis. For all the studied BCGs isophotes appear to be co-centered. Position angles are measured from N through E.

*A2335.* Due to the distance of this galaxy ($z \sim 0.17$) the deviations of the isophotes from perfect ellipses are poorly determined, as showed by the large differences between computed deviations. The galaxy presents a significant change of the PA (40 degree per decade in radius) and an increase of the ellipticity (differences at small radii are due to the different seeing conditions between the exposures).

*A2405.* The parameters of the isophote shapes in the two filters agree well with each other. The ellipticity increases from 0.1 to 0.15. The position angle of the semimajor axis of this galaxy is slowly decreasing at a rate of 15 degrees per decade in radius from 1 to 10 arcsec, then it shows a sharp increase in both filters. At the same radius the isophotes are asymmetric (a3 $\neq$ 0). The asymmetry is probably the reason for the a4 twist.

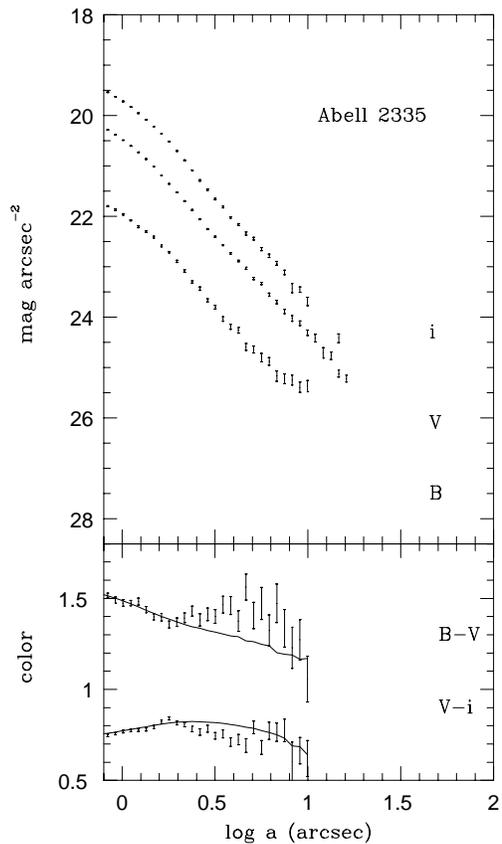
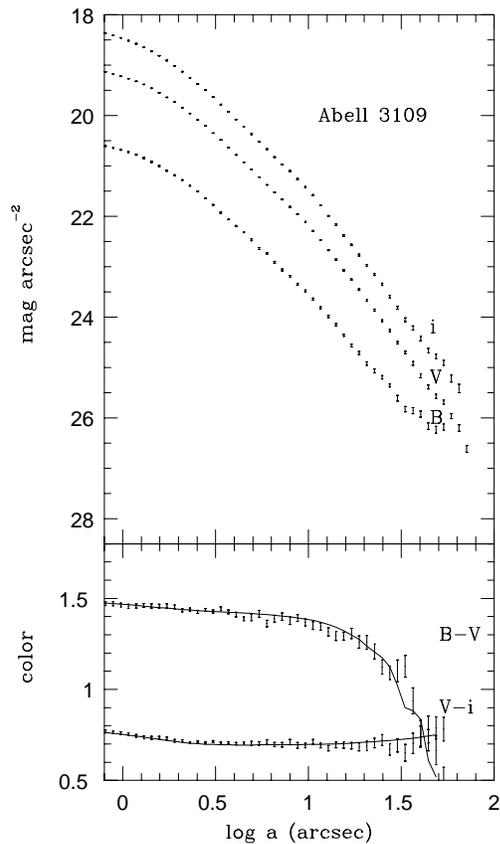
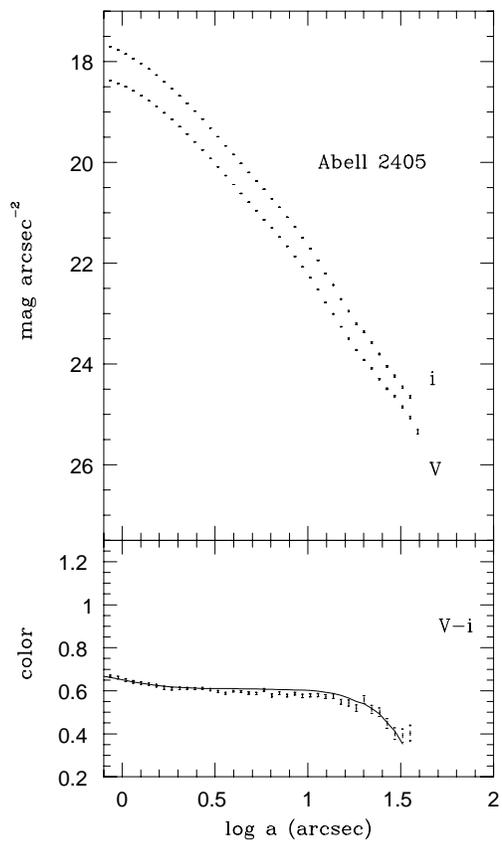
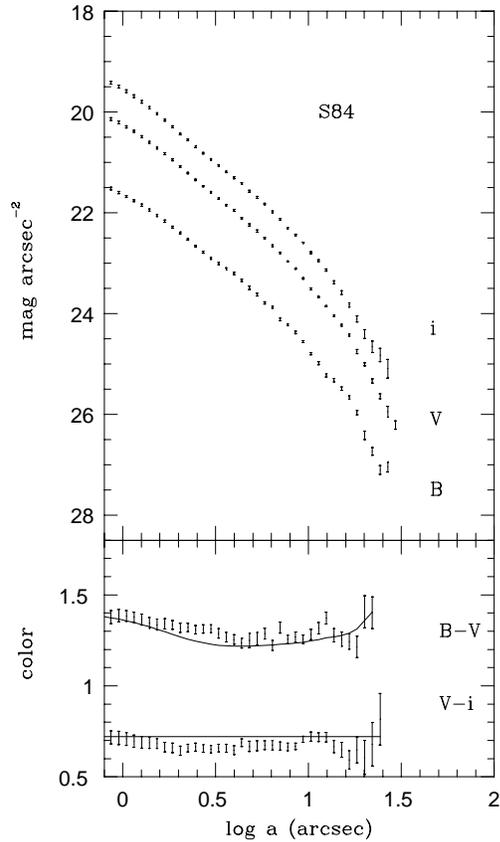

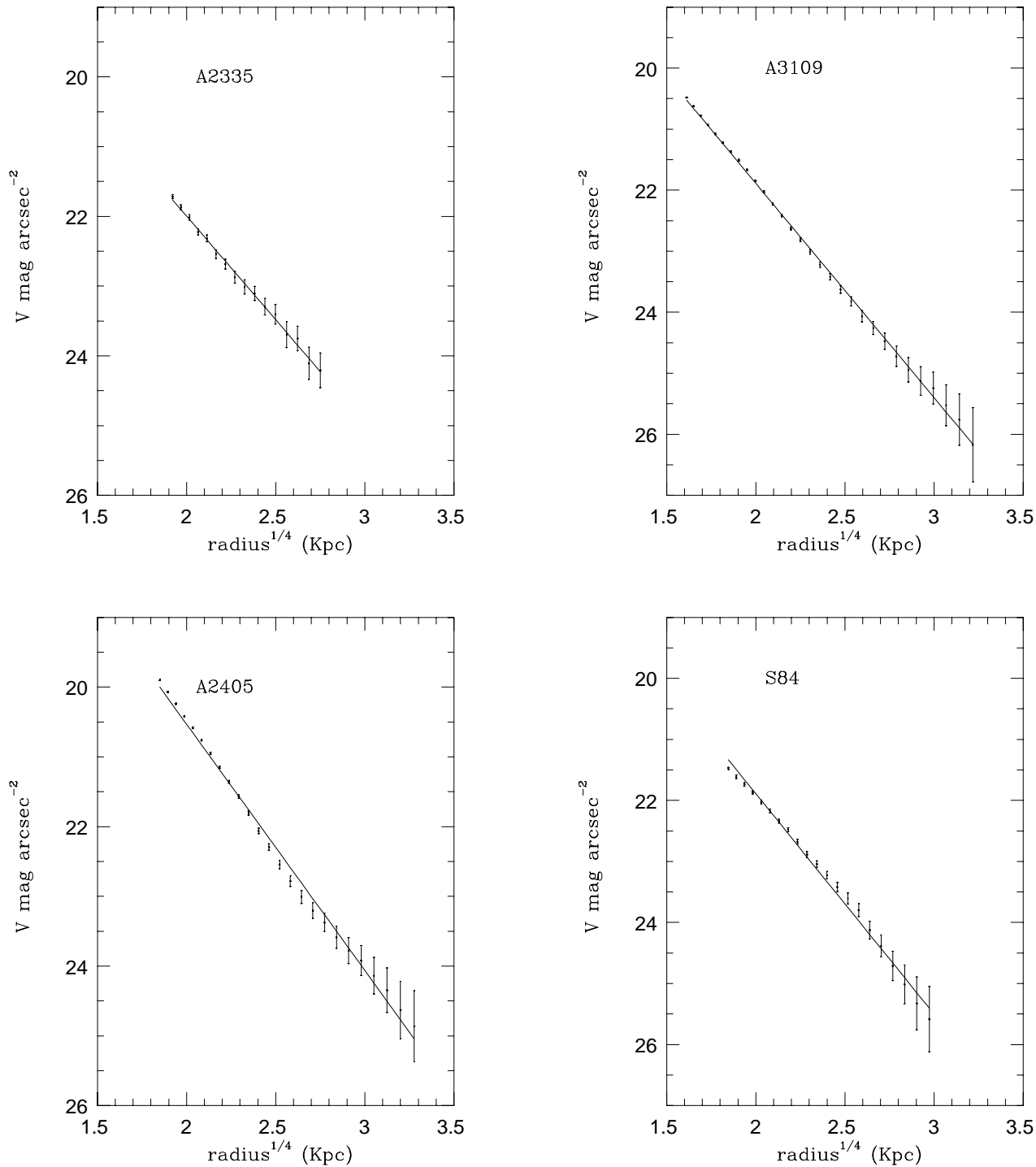

*A3109.* For this BCG, the deviations of the isophote shapes from perfect ellipses are more scattered in the $B$ filter than in the $V$ and $i$ filters. Furthermore, it seems that the $B$ profile is affected by the presence of a galaxy located at a distance of $\sim$ 6 arcsec to the NW, probably because of the lower S/N ratio of the $B$ observations and the blue color of this object

**Fig. 3.** V surface brightness profiles of A2335, A2405, A3109 and S84 plotted as a function of $r = (\sqrt{ab})^{1/4}$. The errors shown are the combination of counting statistics and systematic errors. The solid lines are the best fitting de Vaucouleurs' laws.

(1 mag arcsec$^{-2}$ bluer than the BCG). No clear deviations of the isophotes from perfect elliptical shape have been detected. Ellipticity and PA profiles, as measured in the three filters, agree with one another. The ellipticity increases from 0.2 to 0.32 at a rate of 0.15 per decade in radius. The position angle of the semimajor axis of this galaxy is constant at 20 degrees.

## 4.2. Surface brightness profiles of BCGs in the new sample

Surface brightness profiles of the BCGs were obtained in each filter by integrating the counts within concentric ellipses having constant position angle and ellipticity. Superposed stars, galaxies and chip defects have been masked out. An iterative sigma-clipping technique removing 10 to 30% of the brighter pixels has been used (see Sect. 2.4). Figure 2 shows the surface brightness profiles of the 4 BCGs as a function of the logarithm of the semi-major axis, their observed color profiles and the derived profiles expected from the observing conditions.

$V$ magnitude surface brightness profiles have been fitted with an $r^{1/4}$ law between an angular radius corresponding to 3.2 FWHM of the seeing disk and the radius where the galaxy surface brightness reaches 1 % of the sky level. The resulting best fit de Vaucouleurs' law is plotted in Figure 3. The surface brightness profiles of A2335 and A3109 are well described by a de Vaucouleurs' law; the surface brightness profile of A2405 differs from a de Vaucouleurs' law by less than 10 % over two decades in brightness. Some systematic deviations from de Vaucouleurs' law are detected for this galaxy at 8 kpc (a small excess) and at $\sim$ 40 kpc (a deficit). At the same radii, the surface brightness profile of S84 shows small deviations, but opposite in sign.

Table 3 lists the effective radius, the effective brightness of all BCGs of our full sample having de Vaucouleurs' profile as well as the richness of the cluster in which they are found. The effective brightnesses of all but two galaxies are within half a magnitude. The lower brightness and the larger effective radii of A2335 classify it as a D type BCG. The effective brightness of A2405 is 1 magnitude brighter for its effective radius. This discrepancy cannot be reduced by assigning to this BCG a redshift lower than the estimated one, nor assuming a different value for the Galactic extinction.

**Table 3.** de Vaucouleurs' parameters of BCGs and cluster richness

| BCG   | $r_e$ kpc | $\mu_e$ $m_V$ arsec$^{-2}$ | R |
|-------|-----------|----------------------------|---|
| A1996 | 30.3      | 23.2                       | 0 |
| A2328 | 29.2      | 22.9                       | 0 |
| A2335 | 61.8      | 24.4                       | 2 |
| A2405 | 30.9      | 21.8                       | 0 |
| A3109 | 31.5      | 23.2                       | 0 |
| A3429 | 30.1      | 23.3                       | 0 |
| A3670 | 44.2      | 23.4                       | 2 |
| S84   | 28.2      | 22.9                       | 0 |

Note: The effective surface brightness of A3109 has been corrected using as Galactic extinction the value listed for other BCGs of similar Galactic latitude because the Stark et al. catalogue (distributed data) has no entry for this southern declination.

## 4.3. Color profiles for BCGs in the new and old sample

### 4.3.1. New sample

*A2335.* At small galaxy radii ($r \leq 2$ arcsec) the color profiles of this galaxy closely follow the color profile predicted from the different seeing conditions (Figure 2). Between 2 and 5 arcsec the $V-i$ color profile decreases faster than what would be expected because of an error in the sky level, so that A2335 presents a blue color gradient at intermediate radii. The observed $B-V$ color profile presents a bump at intermediate radii ($\sim$ 4 arcsec) that cannot be caused by an error in the sky determination because the latter produces only monotonical gradients. This does not necessarily mean that this galaxy presents an intrinsic bump in the $B-V$ color, since our procedure to detect color gradients in the outer regions of the galaxies assumes that no color gradients are present. If this is not the case the procedure is just able to detect the presence and the sign of the color gradient, but not its exact shape. So, at intermediate radii A2335 is intrinsically 0.1 to 0.2 mag arcsec$^{-2}$ redder in the $B-V$ and bluer in the $V-i$ color. The $B-i$ color profile does not show at all this behavior, meaning that the cause of the color gradient is related only to the $V$ profile.

*A2405.* The observed $V-i$ color profile agrees with the expected color profile (Figure 2).

*A3109.* The observed color profiles of this galaxy agree with the expected color profiles (Figure 2).

*S84.* The $V-i$ color profile presents small, but systematic differences with the predicted shape (Figure 2). The $B-V$ color profile is never well described by the expected profiles. Inspection of the image itself and of a model-subtracted image of S84 shows that the BCG brightness is contaminated up to the galaxy center by a galaxy located 4 arcsec Eastward and, partially, by a bright star located 7 arcsec Northward.

### 4.3.2. Old sample

*A1644.* This galaxy is not modeled at all by a de Vaucouleurs' law (see Paper I). Consequently the procedure used to predict the inner color gradient is useless. Outer color profiles are flat and therefore the procedure used to predict outer color gradients does not bring any further information to the previous finding, no color gradients are present in the whole studied galaxy radial range.

*A1996.* The observed color profiles of this galaxy agree with the expected color profiles (Figure 5).

*A2328.* The $B-V$ color profile of this galaxy does not follow at all the expected color profile, showing a slow and constant blue color gradient larger than the one caused by an incorrect choice of the sky value (Figure 5). The $V-i$ color gradient shows an extended outer region 0.1 mag arcsec$^{-2}$ redder than the center (Figure 5).

*A3429.* Both observed color profiles of this galaxy agree with the expected color profiles (Figure 5). In the $B-V$ color a very small bump at $\log(a) \sim 0.5$ is present.

*A3670.* The observed $V-i$ color profile of this galaxy agrees with its expected color profile (Figure 5). The $B-V$ color profile shows a constant bluing trend which cannot be accounted by an error in the sky determination (Figure 5).

## 5. Discussion & conclusions

### 5.1. Are visually classified cDs true cDs?

The four new BCGs presented in this paper have been selected because in the ACO catalogue they are visually classified as cDs or are noted to be the dominant galaxy of a B-M I cluster (i.e. satisfying the Morgan, Kaiser, White (1975) cD definition). Among the five BCGs presented in Paper I, one (A1644) is classified in the ACO catalogue as cD, two (A2328 and A3429) are noted to have an outer halo (and they are in B-M I clusters) whereas the remaining two (A1996 and A3670) are not classified (nevertheless they are the dominant galaxies in B-M I clusters). For all these BCGs, but one, we measured the surface brightness profiles down to 27, 26, 25 mag arcsec$^{-2}$ in the $B$, $V$, $i$ filters respectively, and no halo has been detected. Following Schombert's (1988) or Tonry's (1987) quantitative definitions, these galaxies are not cDs. Therefore, the cD ACO class does not contain only true cDs and must be considered as a list of *possible* cDs. The same probably applies to the other lists of BCGs visually classified as cDs.

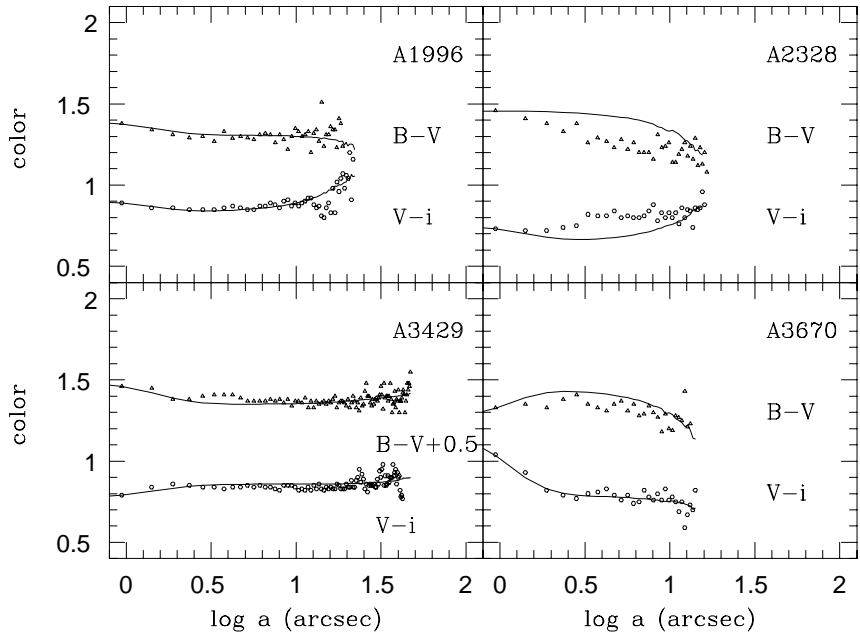

**Fig. 5.** Color profiles for the 5 BCGs of our old sample as a function of the logarithm of the semi-major axis $a$. The lines plotted represent the predicted color profile due to different seeing conditions and errors in the determination of the sky value.

*5.2. Correlation between BCG morphology and cluster richness*

Let us now consider the effective radius of the BCGs as a measure of the slope of their surface brightness profiles. The surface brightness profile of A1644 is shallower than the others that have $r_e \sim 30$ kpc (Paper I) so that we can assign it a much larger $r_e$ (no matter what the exact value is). All the 6 BCGs in richness 0 clusters (A1996, A2328, A2405 A3109, A3429 and S84) have similar effective radii of $\sim 30$ kpc. All the 3 BCGs in richness 2 clusters (A1644, A2335 and A3670) have significantly larger effective radii or equivalently shallower profiles (see Table 3). So, in our sample, BCGs in richer cluster have shallower surface brightness profiles.

We tried to enlarge the BCG sample with data from the literature, mainly to verify that the correlation we found is not the result of some selection bias. Data from the literature are mainly based on BCGs in northern clusters whose richness is estimated by Abell (Abell, 1958) and not on BCGs in the southern clusters whose richness is estimated by Corwin and Olowin on the basis of a somewhat different criterion (Abell, Corwin and Olowin (1989)). Since the richness estimates of the two catalogues for clusters in the zone of overlap do not agree (more than 50 % of the clusters have different richnesses, as noted by Scaramella et al. (1991)), we cannot merge our southern sample with the whole sample from the literature.

Among the clusters whose richness is estimated by Corwin and Olowin, as the ours, we find 15 other BCGs with published effective radii. Table 4 lists these BCGs with their effective radius, the richness of the cluster in which they are found and the reference to the literature.

The BCGs A2328 and A1644 (not listed in Table 4) have also been previously observed (Hoessel & Schneider 1985). For A1644 the effective radius quoted by these authors (74 Kpc) is consistent with our determinations ($r_e$ larger than 30 Kpc), considering that the surface brightness profile of this galaxy does not obey to the de Vaucouleurs' law. The effective radius quoted for the BCG A2328 (7 Kpc) is not consistent with our determination and seems too small for the BCG luminosity and, in general, for BCGs (for example in Schombert's (1987) sample no BCG has such as small effective radius). We note that the choice of the exact value for the effective radius of this BCG (less than 30 Kpc) does not alter the correlation we found.

In Fig. 6 we plot the effective radius vs. cluster richness for all the BCGs in our sample and in Table 4. The correlation is clear, although with a large dispersion in the effective radius. Within the sample from the literature, only the BCGs A2877 and A2384 deviate somewhat from this correlation, possessing an effective radius which is larger than expected for the cluster richness.

What are the implications of such a correlation for BCG formation? There are two main scenarios explaining BCGs properties. In the first scenario, the BCGs morphology is fixed by the initial conditions. In this case, the effective radius-cluster richness correlation constrains the initial conditions: they must be such as to form BCGs with shallow profiles in rich clusters. In the second scenario, the BCG environment modifies the morphology of BCGs, and, as a consequence, it contemporaneously evolves. In this scenario the correlation poses some constraints on the mechanisms that can form the BCGs. Since mergings are good candidates for BCG formation and are more probable in rich clusters (all other cluster characteristics fixed), the correlation we found tells us that the merger products tend to have shallower profiles, as N-body simulations (starting with Aarseth & Fall 1980) also indicate.

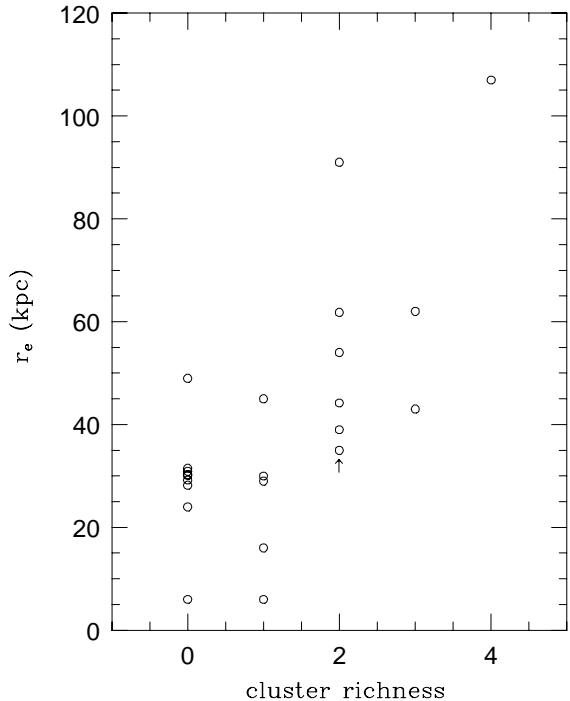

**Fig. 6.** Effective radius of BCGs as a function of the richness of the cluster in which they belong. The arrow marks the lower limit of the effective radius of the BCG A1644. The absence of small BCGs ($r_e < 30$ kpc) in rich clusters ($R \geq 2$) and the scarcity of large BCGs ($r_e > 30$ kpc) in poor clusters ($R < 2$) are noteworthy.

**Table 4.** Cluster richnesses and BCG effective radii in the literature

| Cluster | R | $r_e$ kpc | reference |
|---|---|---|---|
| A22 | 3 | 62 | Hoessel & Schneider 1985 |
| A42 | 2 | 91 | Hoessel & Schneider 1985 |
|  |  | 91 | Schombert 1987 |
| A140 | 1 | 29 | Hoessel & Schneider 1985 |
| A389 | 3 | 43 | Hoessel & Schneider 1985 |
| A514 | 1 | 16 | Schombert 1987 |
| A2347 | 0 | 6 | Hoessel & Schneider 1985 |
| A2384 | 1 | 45 | Hoessel & Schneider 1985 |
| A2462 | 0 | 24 | Hoessel & Schneider 1985 |
| A2554 | 2 | 35 | Hoessel & Schneider 1985 |
| A2686 | 1 | 6 | Hoessel & Schneider 1985 |
| A2877 | 0 | 49 | Schombert 1987 |
| A3159 | 2 | 39 | Schombert 1987 |
| A3186 | 1 | 30 | Boldrini 1989 |
| A3558 | 4 | 107 | Schombert 1987 |
| A3639 | 2 | 54 | Garilli, Maccagni, Vettolani 1991 |

### 5.3. Color and color gradients

The nuclear colors of all BCGs are the same in the rest-frame, the color dispersion being comparable to the photometric accuracy (see Table 5).

The procedure presented in Sect. 3.2 allows us to detect color gradients of smaller amplitude (0.01 mag arcsec$^{-2}$ per decade in radius) and over larger galaxy radial ranges than previously, disentangling color gradients intrinsic to the BCG from predicted effects. Table 6 summarizes the results. The suspected (Paper I) color gradients in the BCGs A2328 and A3670 are confirmed to be intrinsic to the galaxies. A color gradient is detected in the outer region of the BCG A2335 whereas the one observed in S84 appears to be predicted. For all the other BCGs no intrinsic color gradients are present.

Only in one case can our color profiles be compared with other authors', i.e. in the case of A1644. McNamara and O'Connell (1992, hereafter MO) do not find any significant color gradient in the $b-I$ color, as we do not in the $B-i$ color. It has to be noted that the comparison can be made only for the bright part of the profile because of the apparent size of the galaxy. The bluer nuclear region, suspected by MO, is not confirmed by our higher resolution images.

**Table 5.** BCG nuclear colors[a] in the rest-frame

| BCG | B-V | V-i | E(B-V)[b] |
|---|---|---|---|
| A1644 | 0.95 | 0.62 | 0.026 |
| A1996 | 0.84 | 0.80 | 0.066 |
| A2328 | 0.74 | 0.71 | 0.025 |
| A2335 | 0.78 | 0.76 | 0.026 |
| A2405 | – | 0.56 | 0.015 |
| A3109 | 1.21 | 0.67 | 0.02[c] |
| A3429 | 0.66 | 0.81 | 0.025 |
| A3670 | 0.74 | 0.56 | 0.050 |
| S84 | – | 0.67 | 0.009 |
|  | $0.85 \pm 0.18$ | $0.68 \pm 0.09$ |  |

[a] The correction for extinction in our Galaxy has not been applied since it is negligible compared to the photometric error.
[b] The E(B-V) has been calculated following Schneider et al. (1983) from the $N_H$ column density of Stark et al. (distributed data).
[c] Estimated (see the note of Tab. 3)

**Table 6.** BCGs intrinsic color gradients

| BCG | B-V | | V-i | | R |
|---|---|---|---|---|---|
|  | inner | outer | inner | outer |  |
| A1644 | – | no | – | no | 2 |
| A1996 | no | no | no | no | 0 |
| A2328 | blue | blue | – | red | 0 |
| A2335 | no | red | no | blue | 2 |
| A2405 | – | – | no | no | 0 |
| A3109 | no | no | no | no | 0 |
| A3429 | no | no | no | no | 0 |
| A3670 | blue | blue | no | no | 2 |
| S84 | – | – | – | – | 0 |

Color gradients of galaxies in our sample are compatible with the small BCG color gradients measured by MVC but not with the values quoted by MO. In fact in our sample 4 BCGs out of 7 do not show any color gradient in the $B-V$ and 6 out of 8 do not show any in the $V-i$. In the MVC sample 11 BCGs out of 14 in the $B-V$ and 10 BCGs out of 14 in the $V-I$ possess a color gradient less or equal than 0.10 mag per decade in radius. Only $\sim 15\%$ (3 BCGs out 19) of the MO sample show a gradient compatible with 0 whereas half of their sample shows a color gradient larger 0.15 mag per decade in radius. The MO sample thus contains a higher proportion of galaxies with strong color gradients than the MVC and the present samples. Several reasons can be brought up to explain such a discrepancy. First of all, the three samples contain a small number of galaxies, and infering overall properties of a class from small sized samples should be done with care. More even so in this case, where color gradients could be produced by different seeing conditions between the exposures and by errors in the determination of the sky value (and sky determination can be extremely tricky in images where the object covers a large part of the field of view).

On the other hand, the two samples could just be different, since the selection criterion was different. We selected galaxies in Bautz-Morgan type I clusters, while MO based their inclusion on the presence of cooling flows. Pesce et al. (1990) have shown that more than 75% of the clusters should possess a cooling flow, and this would imply that some superposition exists in the two samples; furthermore, MVC showed that the presence and the amplitude of a cooling flow are not clearly related to the BCG color gradients. Therefore the two criteria seem to select similar BCGs.

Probably, the best way to clear up the effect of the environnement on BCGs is to obtain color profiles for larger samples of BCGs.

*5.4. Other correlations*

In our sample, we do not see any relation between BCG color gradient and cluster richness (but we span only two richness classes), and between color gradient and BCG effective radius. Moreover, we do not see any correlation between color gradient and BCG morphology (gE or D) either in MVC and our samples. Finally, color gradients of MVC BGCs are not related to the B-M type of the cluster.

What does the absence of correlations between color gradients of BCGs and their environment imply?

Mergers tend to modify color gradients. If the core of the cannibalized galaxy survives for a long time in the host BCG (Navarro, 1989), mergers tend to produce a new color gradient; in the opposite case mergers tend to erase an already existing color gradient (White 1980). Tidal stripping (Gallagher and Ostriker 1972; Hausman and Ostriker 1978) would produce blue color gradients because the color-magnitude (Visvanathan and Sandage 1977) and the color-radius relations (Sandage and Visvanathan 1978) imply a bluer color for the stripped material accreting on the BCGs. Cooling flows, if converted into stars, would produce color gradients since the new stars are younger and probably have spatial distributions and metallicities different from the older BCG stars. More generally, it is easy to convince oneself that any stellar population different in age, metallicity or mass function from the BCG one, induces a variation in the shape of the color profile at least if its spatial distribution is not a fine tuned function of the BCG one.

The absence of any correlation between color gradient and characteristics of the BCG environment argues against a strong and recent influence of the environment on the BCG morphology. If we also take the different morphologies of BCGs as indicative of different environmental influences, this conclusion is supported by the absence of correlation between BCG morphology and color gradients. This conclusion can be also extended to the cD envelopes since their colors are not significantly different from the main body of the BCGs (Mackie 1992). However, this does not exclude that for *some* BCGs the environment is playing a fundamental role as is probably the case for GREG (Maccagni et al. 1988) or for the nine nuclei A407 (Mackie 1992) or maybe for our BCGs with color gradients.

We cannot set a limit to the role played by the environment during the early life of the BCG as long as we cannot be sure that environmental properties do not change during cluster evolution. But we exclude a strong influence at the present time in agreement with Merritt's (1984) statement that BCG morphology changes very little after cluster formation and collapse.

*Acknowledgements.* SA thank the Director of the Istituto di Fisica Cosmica of the Consiglio Nazionale delle Ricerche, Prof G. Boella, for the hospitality in his Institute where most of the work was done. We thank A. Cappi that observed the galaxies analysed in this paper and S. Pocar for his help in convolving BCGs models with seeing. We thank the referee, Dr. G. Mackie, and Dr. E. Davoust for their numerous suggestions improving the style of this paper. SA was partially supported by a fellowship of the Trento University.